\documentclass[twoside,11pt]{article}

\usepackage{jmlr2e}
\usepackage{graphicx}
\usepackage{subfigure}
\usepackage{dirtytalk}
\usepackage{longtable}
\usepackage{multirow}
\usepackage{tabulary}
\usepackage{multirow}
\usepackage{tabulary}
\usepackage{placeins}
\usepackage[ruled]{algorithm2e}
\usepackage{algpseudocode}

\firstpageno{1}

\begin{document}

\title{Predicting Higher Education Throughput in South Africa Using a Tree-Based Ensemble Technique}

\author{\name Rendani Mbuvha \email rendani.mbuvha@wits.ac.za \\
       \addr School of Statistics and Actuarial Science\\
       University of Witwatersrand \\
         Johannesburg, South Africa
       \AND
       \name Patience Zondo \email patiezondo28@gmail.com \\
       \addr   School of Statistics and Actuarial Science\\
       University of Witwatersrand \\
         Johannesburg, South Africa
        \AND
       \name Aluwani Mauda  \email  aluwanimauda@gmail.com \\
       \addr  Moshal Scholarship Program\\
       Johannesburg, South Africa 
    \AND
       \name Tshilidzi Marwala \email  tmarwala@uj.ac.za \\
       \addr  Office of the Vice-Chancellor, University of Johannesburg \\
           Johannesburg, South Africa
       }

\editor{}

\maketitle

\begin{abstract}
We use gradient boosting machines and logistic regression to predict academic throughput at a South African university. The results highlight the significant influence of socio-economic factors and field of study as predictors of throughput. We further find that socio-economic factors become less of a predictor relative to the field of study as the time to completion increases. We provide recommendations on interventions to counteract the identified effects, which include academic, psychosocial and financial support. 

\end{abstract}

\section*{Introduction}
\label{intro}

Efficient higher education systems are central to the developmental agenda of any nation. Such systems are mainly characterised by high progression and throughput rates at both the undergraduate and postgraduate levels   \citep{montmarquette2001determinants}. Student success indicators are therefore of interest to governments, higher education policymakers and funders. This interest is not only driven by  demands of the labour market, but also for accountability to taxpayers and donors \citep{yorke2004retention}.

Throughput rates in South African universities have remained alarmingly low. Given South Africa's history, this is especially true for Black African students, who constitute the majority of higher education enrollment, growing from 61\% of total enrollment in 2005 to 73\% in 2017 \citep{che2019}. According to the National Plan for Higher Education (NPHE) compiled by the South African Department of Education in 2001, the reported graduation rate of 15\% in South Africa was amongst the lowest in the world \citep{departmentofeducation2001education}. In the 2005 NPHE, the Department of Education states that out of the 120 000 students who enrolled in higher education in the year 2000, 36 000 (30\%) dropped out in their first year of study \citep{departmentofeducation2005education}. A further 24 000 (20\%) dropped out during their second and third years. Of the remaining 60 000, (22\%) graduated within the specified three-year duration for a generic Bachelors degree \citep{departmentofeducation2005education}. Subsequently, the department issued a public statement stating that the drop-out rate was
costing the national fiscus USD 650 million in grants and subsidies to higher education institutions without a proportionate return on investment. A decade later, the statistics show little improvement.
The 2017 Vital Stats report by the Council on Higher Education reports that the throughput rate for the 2012 cohort is 29\% for students enrolled in a 3 year degree\citep{che2019}. Moreover,a panel study of the 2009 cohort shows that the throughput rate after 10 years is still only 60\%, highlighting the gross inefficiency of the system \citep{dhet2020}.

In this work, we develop gradient boosting machines and logistic regression models for predicting university wide throughput based on socio-demographic variables and enrolled programme characteristics. We further investigate the relative variable importance of each of the variables using gradient boosting machines. Our definition of throughput is the proportion of students within a cohort who complete their degree in the minimum recommended time. This work makes a significant contribution to the application of machine learning techniques on South African and African higher education data where literature in this area has been limited. Further contributions include a presentation of variables that are found to be important in predicting throughput, which can then serve as policy signals for interventions in this sector.

	 
\section*{Related Work}

There have been numerous studies in modelling higher education academic throughput. The Canadian Educational Policy Institute \citep{epi2008access} analyses a range of literature to outline the factors that affect post-secondary access, persistence and barriers. They find important factors including inter alia; gender, disability, parental education and finances \citep{epi2008access}.  

\cite{van2017higher} investigate university access, throughput, and dropout by tracking the matriculation class of 2008 from 2009 to 2015 across all South African universities. The study finds that the students who get admitted to university often do not complete their degrees on time and a significant number of them never get to complete their degrees at all \citep{van2017higher}. In the South African context \cite{van2017higher} find that high-school academic achievement is highly predictive of university access and also of university success, though to a lesser extent. \cite{geiser2007validity} investigate if high-school grades and standardized test scores predict college progression outcomes in the long term and short term. The study finds that high-school grade point average (HSGPA) is the best predictor not only of freshman grades in college, but of four-year college outcomes as well. 

\cite{duff2004understanding} explores the relationships between learning approaches, age, gender, prior academic achievement and their subsequent progression. \cite{duff2004understanding} finds, similar to \cite{geiser2007validity}, that the most significant predictor of first-year academic performance and progression is prior academic achievement (i.e. performance in high school examinations).  \cite{duff2004understanding} also states that prior academic success was positively associated with critical engagement with learning material and academic self-confidence, and inversely related to rote-learning approaches and lack of direction.

To explain how socioeconomic status plays a role in academic progression \cite{perry2015analysis} examines the various federal, state, institutional, employment, and loan programs accessible to students at an Illinois community college to consider the effect of these programs on student graduation rates and grade point averages. The study concludes that there is a statistically  significant difference in graduation rates of students who received grants, institutional scholarships, and federal work study over students who received only loans or no financial assistance \citep{perry2015analysis}. Similarly \cite{cabrera1992role} empirically studies the role of finances on college persistence by presenting a causal model that relies on numerous theoretical frameworks. Their results suggest that financial aid, and its associated benefits, are important because they balance opportunities between well-off and low-income students \citep{cabrera1992role}.

\cite{spady1970dropouts} suggests a sociological model of the dropout process and concludes that the decision to drop out is as a result of a complex social process.  \cite{montmarquette2001determinants} view the problem of persistence in university as an interaction between an experiment in school and an experiment in the labour market finding that students who pursued both the labour market and university education had a higher chance of dropping out. \cite{montmarquette2001determinants} also propose examining part time student throughput rates vs that of full time students. 


The different authors discussed above all use various statistical methods in their analysis. A multiple regression was utilised by \cite{spady1970dropouts} to analyse how the factors independently explained the reason for dropout. \cite{geiser2007validity}  use ordinary linear regression to study the relationship between admissions factors and cumulative fourth-year grades, while logistic regression is employed in the analysis of four-year graduation. \cite{montmarquette2001determinants} use a bivariate probit model to model the decision of dropping out or persisting in university.  \cite{strugnell2017throughput} employ survival analysis in investigating the  impact of socio-demographic and high school attainment on student throughput in the actuarial science programme at the University of Cape Town. \cite{strugnell2017throughput} find that race, financial and psychosocial support, as well as national benchmark tests are strong predictors of success in the programme. 

 \cite{musso2020} use Artificial Neural Networks (ANNs) to predict key elements of a student's trajectory at university. Their findings show a significant level of accuracy in this task while also showing that learning and coping strategies, as with \cite{duff2004understanding}, are found to be the most important predictors of grade point average and degree completion. Numerous other studies employ ANNs in the prediction of various higher education student outcomes \citep{musso2013predicting,lau,ajams2015343,article_Abu-Naser}. 

Other machine learning approaches include \cite{article_Kov} who investigate socio-economic and study environmental factors that affect dropouts using Classification and Regression Trees (CART). \cite{ramaswami2010chaid} use a Chi-square automatic interaction detection (CHIAD) decision tree to extract grade prediction rules in higher secondary schooling in India. \cite{aulck2017stem} use logistic regression, gradient boosting and random forest to predict dropout rates from STEM programmes at the University of Washington. \cite{huang2011predictive} use different data mining techniques such as multiple linear regression (MLR), multilayer perceptron (MLP) networks, radial basis function (RBF) networks, and support vector machines (SVM) to academic performance of students in an engineering course. \cite{huang2011predictive} came to a conclusion that  that radial basis function (RBF) networks, and support vector machines perform equally well. All these models had an accuracy that was greater than 80\%.

\cite{van2017higher} use a multivariate analysis to analyse the response variables: access, throughput and retention. \cite{duff2004understanding} uses linear regression and cluster analysis to conclude on which learning approaches are effective as well as the significant variables in the prediction of performance.

\cite{perry2015analysis} uses a combination of a t-test and Chi-squared test for the analysis of the differences that exist in graduation rates and grade point averages of students with any type of financial aid compared to students with no financial aid. 


We take a similar approach to \cite{huang2011predictive} and use machine learning techniques to predict progress outcomes in higher education. However we employ gradient boosting machines to predict throughput and degree completion. This is different to \cite{huang2011predictive} because instead of focusing on one course this work focuses on making predictions on a university wide dataset. Gradient boosting machines, which have become prominent in data mining, are used as they achieve great predictive accuracy in real world applications that utilise tabular data, as is the case in this study \citep{zhou2012ensemble}. The rationale presented by \cite{marsland2015machine} for using ensemble techniques is that when multiple learning machines (decision trees in our case) are combined to predict the same response variable they are likely to give higher accuracy results than one learning machine. To test if this argument holds, a logistic regression model is used as a benchmark.


\section*{Methods}\label{methods}

\subsection*{Data}
The data used in this study consists of enrolment and degree completion information from a large university in South Africa for the period 2007-2013. The data includes socio-demographic information at the point of first enrollment for students enrolling for 3 and 4 year degrees across the university.

The data includes 36286 unique student observations with 6 explanatory variables. We split the data such that  80\% is used for model training, 10\% for model validation and the remaining 10\% for testing. The variables considered in the model building are listed in table \ref{explanatory}.

The prediction target variable of interest is the binary indicator of whether a student completes their degree in the minimum required time, N years. We additionally also  consider predicting completion in N+1 and N+2 years. 
 
The authors who handled the data used in this work were required to sign \say{preservation of confidentiality of students' information} agreements and as such unique identifiers to individual students were anonymised.  
\begin{table}[t]
\caption{Description of variables used for  modelling degree completion}
\resizebox{\linewidth}{!}{
\begin{tabular}{ll}
\hline
\textbf{Variable Name} &\textbf{Description} \\
\hline
 Year of First Enrollment & Specifies the calendar year of first enrollment at the university \\
 Year of Study & Specifies the year of study the particular record is in \\
 Degree Duration & The duration of the degree registered for\\
 Field of Study & The field of study under which the student is registered\\
Age & The student's age at first enrollment \\ 
Race Description & Race description of the student according to national classification \\  
Sex & Specification of the sex of a particular student \\  
N Completion & Binary indicator of whether the student completed their degree in the minimum require N years\\
N+1 Completion & Binary indicator of whether the student completed their degree in N+1 years\\
N+2 Completion & Binary indicator of whether the student completed their degree in N+2 years\\
\hline\noalign{\smallskip}
\end{tabular}}
\label{explanatory}
\end{table}

\subsection*{Ensemble Techniques and Gradient Boosting Machines}

\citet{Perrone1992networks} define an ensemble technique as a machine learning method that combines the strengths of a group of base models to produce a prediction model which has a higher predictive performance.

Ensemble methods are further divided into different types such as: bagging, boosting, Bayesian parameter averaging and  Bayesian model combination. Gradient boosting machines are as their name suggests a boosting technique.

Boosting involves combining weak classifiers to create an ensemble classifier with higher performance. Each model is built sequentially taking into account past performance.  Weights are given to each data point according to how often it has been classified correctly in past predictions. At each step a new classifier is trained on the training set, with the weights that are applied to each observation being altered at each step according to how successfully the particular observation has been classified in the past. After each step the miss-classification error is then calculated and the weights for incorrectly classified observations are updated. Weights for correctly classified observations are left unchanged. Training is stopped after a predetermined number of iterations, or after all observations have been correctly classified.

\begin{figure}\centering
	\includegraphics[width=\linewidth]{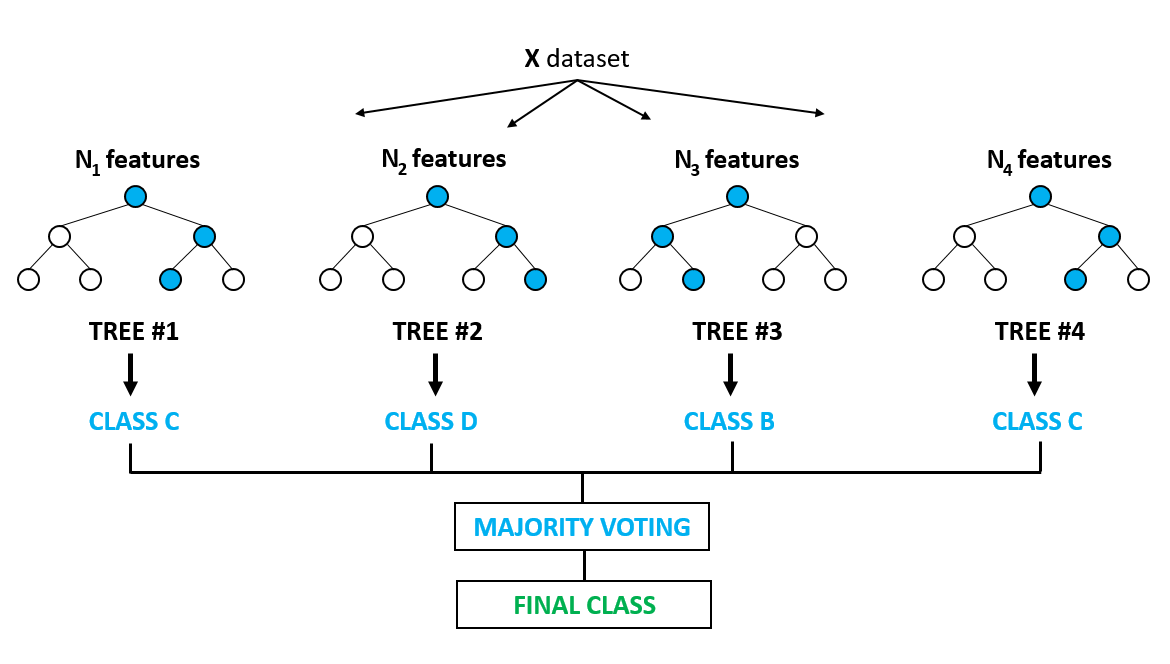}
	\caption{An illustration of the majority vote concept in tree ensemble models} \label{Fig:vote}
\end{figure}




Gradient boosting machines are an example of boosting ensembles  introduced by \cite{friedman2001greedy}. As with many ensemble methods gradient boosting combines outputs of multiple decision trees in order to produce a high performing single predictor that reduces the variance of the individual weak predictors (Author et.al 2019). Gradient boosting machines sequentially  fit  decision  trees in such a way that they learn from what the preceding trees have misclassified  \citep{hastie01statisticallearning}. The sequence of trees fitted thus form a \say{committeee} of models for preforming class predictions on new data points on the basis of a majority vote as illustrated in figure \ref{Fig:vote}\footnote{figure source : http://www.globalsoftwaresupport.com/random-forest-classifier}. 

The sequential fitting of trees on residuals and the majority vote embedded in gradient boosting machines results in a powerful machine learning technique which has proven to be considerably successful in many applications \citep{natekin2013gradient,touzani2018gradient,atkinson2012assessing}. Other advantages of gradient boosting machines over traditional classification models like logistic regression include; their ability to model non-linear relationships between the target variable and inputs, principled handling of missing data and the relaxation of the requirement for pre-processing of data as required by nonlinear models such as neural networks.   

We also employ logistic regression as a baseline model. Logistic regression has  proven robust in a wide range of domains and is usually effective in the estimation of class probabilities for binary variables \citep{long1993comparison}. This type of regression allows for the modelling of the relationship between a binary random variable and a linear combination of predictor variables. The output of the model is a probability of a particular observation belonging to one of the two response categories given the data. In our case the two categories will encompass those who complete and those who do not complete their degrees over a given period of time (N, N+1 and N+2).
\subsection*{Variable Importance in Gradient Boosting Machines}

Variable importance aids in the interpretability of the gradient boosting models. It helps in deciding which variable is important in predicting student outcomes. Predictors are ranked from most important to the least important hence aiding in removing the predictors that are not significant in prediction.
This is accomplished by calculating the relative influence of each variable: whether that feature was selected during splitting when the tree was being built and how much the squared error over all trees decreased as a result of that variable being chosen.

\subsection*{Predictive Performance Measures}
 We use the receiver operating characteristic (ROC) and the precision-recall curves as our main tools for predictive performance assessment. 

 The ROC curve is used when assessing the trade-off between the true positive rate (sensitivity) and the false positive rate (specificity). For a given time to completion, the true positive rate represents the ratio of actual  degree completions in the data that the model correctly identifies as such, to the total number of students that were predicted to have completed. The false positive rate represents the ratio of the number of students that did not complete but were predicted as having completed to the total number of students that were predicted to have completed.
 
 A model's precision is the ratio of the completions predicted correctly to total number of actual completions in the data. Recall is the ratio of the number of passes predicted correctly to the total number of predicted completions. Recall is the same as sensitivity. A perfect classifier is one with a precision close to one as the recall increases and a recall close to one as the precision increasese. The precision-recall curve illustrates this trade off between the precision and recall. In both cases the area under the ROC curve (AUC) and the area under the precision-recall curve (AUPR) give a summary statistic for model performance. In this work we consider AUC and AUPR on a holdout test set as metrics for evaluating predictive performance.

\section*{Results}

In this section we present and discuss the results from the analysis described in the previous section.

\begin{figure*}[htbp]
\centering
   \subfigure[ROC curve for completion in $N$ years ]{
  \includegraphics[width=0.45\textwidth]{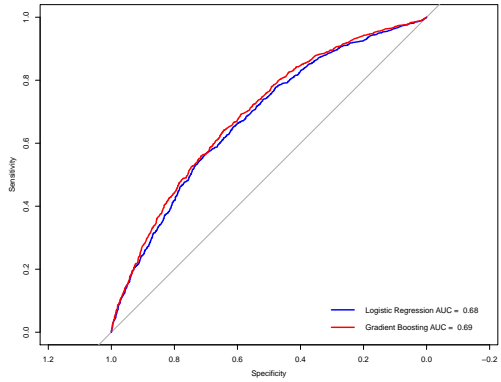}
  \label{fig:auc_n}
  }
  \subfigure[ROC curve for completion in $N+1$ years ]{
  \includegraphics[width=0.45\textwidth]{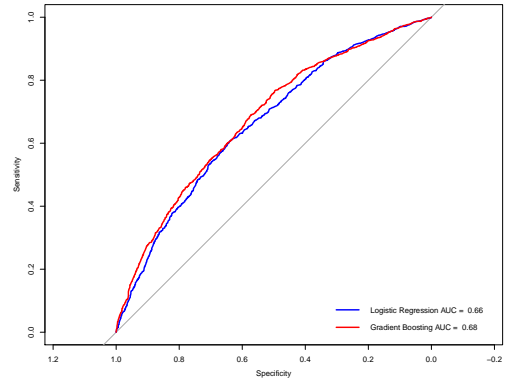}
  \label{fig:auc_n1}}
  \subfigure[ROC curve for completion in $N+2$ years]{
  \includegraphics[width=0.45\textwidth]{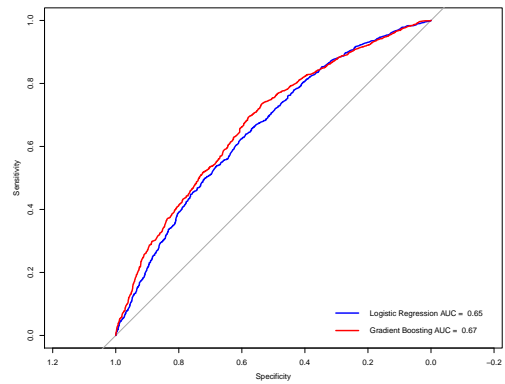}
  \label{fig:auc_n2}
}
 
\caption{Receiver Operating Characteristic curves for the gradient boosting and logistic regression models for  degree in $N$ \ref{fig:auc}\subref{fig:auc_n} , $N+1$ \ref{fig:auc}\subref{fig:auc_n1}, $N+2$ \ref{fig:auc}\subref{fig:auc_n2}. }
\label{fig:auc}
\end{figure*}

Figure \ref{fig:auc} shows the ROC curves for models predicting completion in $N$, $N+1$ and $N+2$ years respectively.  The 45 degree gray line in the figure represents an equal number of true positive rates and false positive rates, if a model is represented by this line, the classifier is as good as a random guess. Curves that are closer to the north west corner of the graph and above the 45 degree gray line represent better classifiers.

When considering models for completion in N years logistic regression results in an AUC  of 0.68, while gradient boosting yields an AUC of 0.69. For completion in N+1 years, the AUC for logistic regression is 0.66 and 0.68 from gradient boosting. Similarly for N+2 completion the AUCs are 0.65 and 0.67 respectively.

\begin{table}[hb]
 \caption{Summary of model predictive performance statistics}
    \centering
\begin{tabular}{c|c|c|c}

 Model  & Degree Completion Time & AUC & AUPR\\
 \hline
Logistic Regression     & N& 0.68 & 0.49\\
Gradient Boosting & N& 0.69 & 0..51\\
Logistic Regression     & N+1 & 0.66 & 0.66\\
Gradient Boosting & N+1 &0.68 & 0.68\\
Logistic Regression     & N+2 & 0.65 & 0.70\\
Gradient Boosting & N+2 & 0.67 & 0.73\\
\hline
\end{tabular}
    \label{tab:perf}
\end{table}



\begin{figure*}[!ht]
\centering
   \subfigure[Precision-Recall curve for  $N$ years ]{
  \includegraphics[width=0.48\textwidth]{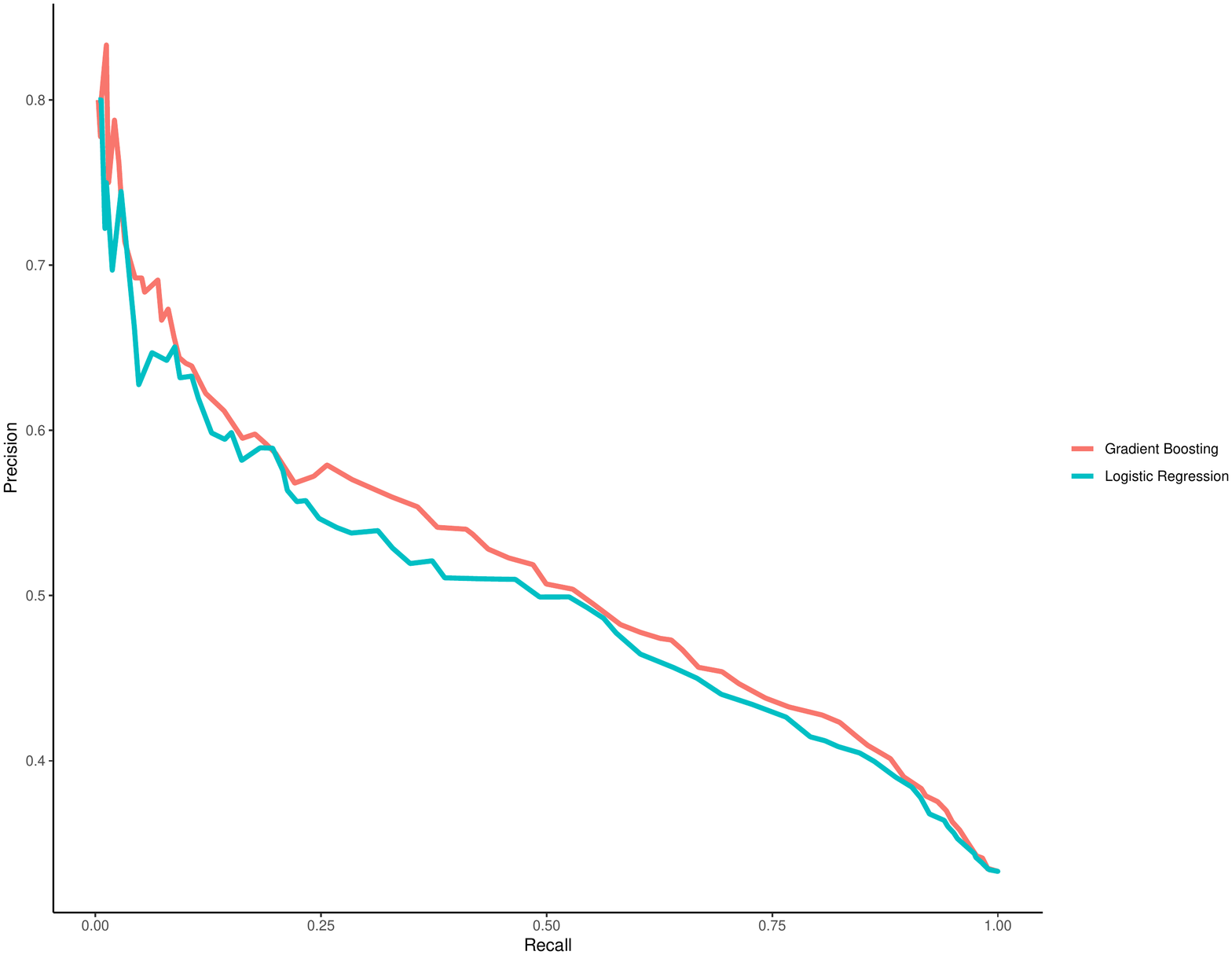}
  \label{fig:pr_n}
  }
  \subfigure[Precision-Recall curve for $N+1$ years ]{
  \includegraphics[width=0.48\textwidth]{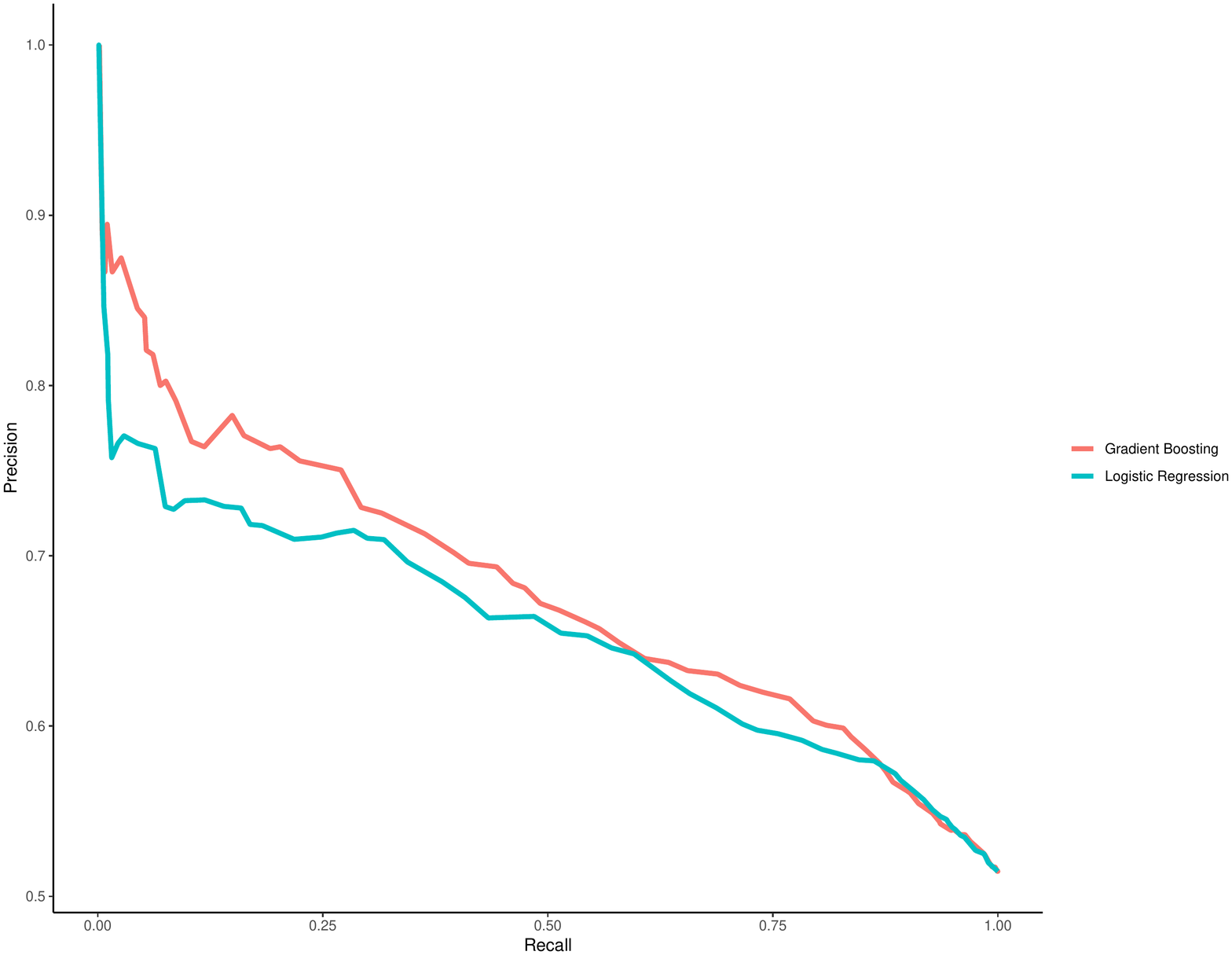}
  \label{fig:pr_n1}}
  \subfigure[Precision-Recall curve for $N+2$ years]{
  \includegraphics[width=0.5\textwidth]{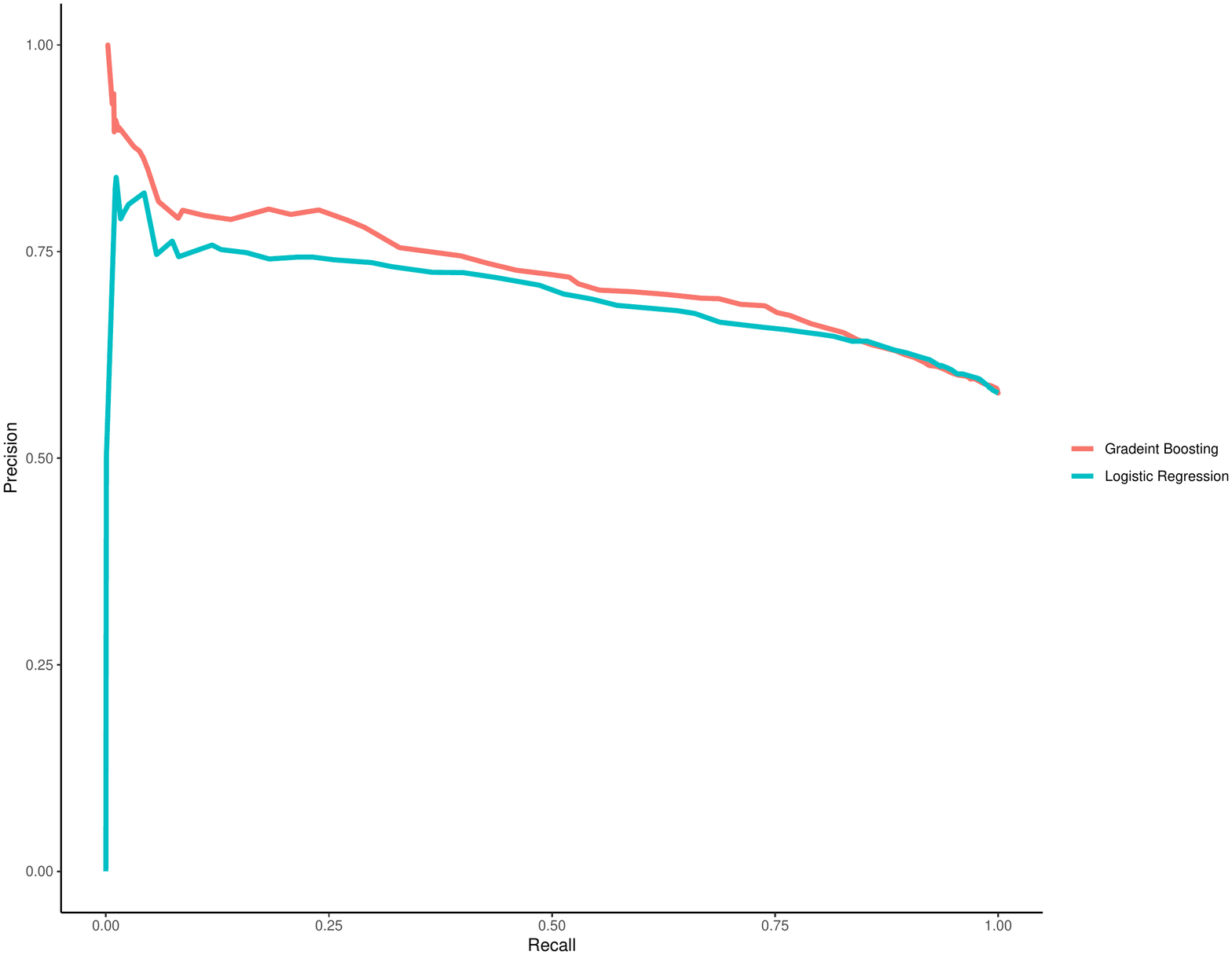}
  \label{fig:pr_n2}
}
 
\caption{Precision-Recall curves for the gradient boosting and logistic regression models for  degree in ($N$)  \ref{fig:pr}\subref{fig:pr_n} , ($N+1$)  \ref{fig:pr}\subref{fig:pr_n1}, ($N+2$)  \ref{fig:pr}\subref{fig:pr_n2}. }
\label{fig:pr}
\end{figure*}

Figure \ref{fig:pr} represents the precision-recall curve  resulting from gradient boosting and logistic regression models discussed above. It can be seen from the the plot that the gradient boosting model displays marginally superior performance relative to logistic regression yielding AUPRs of 0.51 and 0.49 (N), 0.68 and 0.66 (N+1), 0.73 and 0.70 (N+2). Table \ref{tab:perf} shows a summary of the model performance statistics.

\begin{figure*}[htbp]
\centering
   \subfigure[Variable importance for $N$ years]{
  \includegraphics[width=0.47\textwidth]{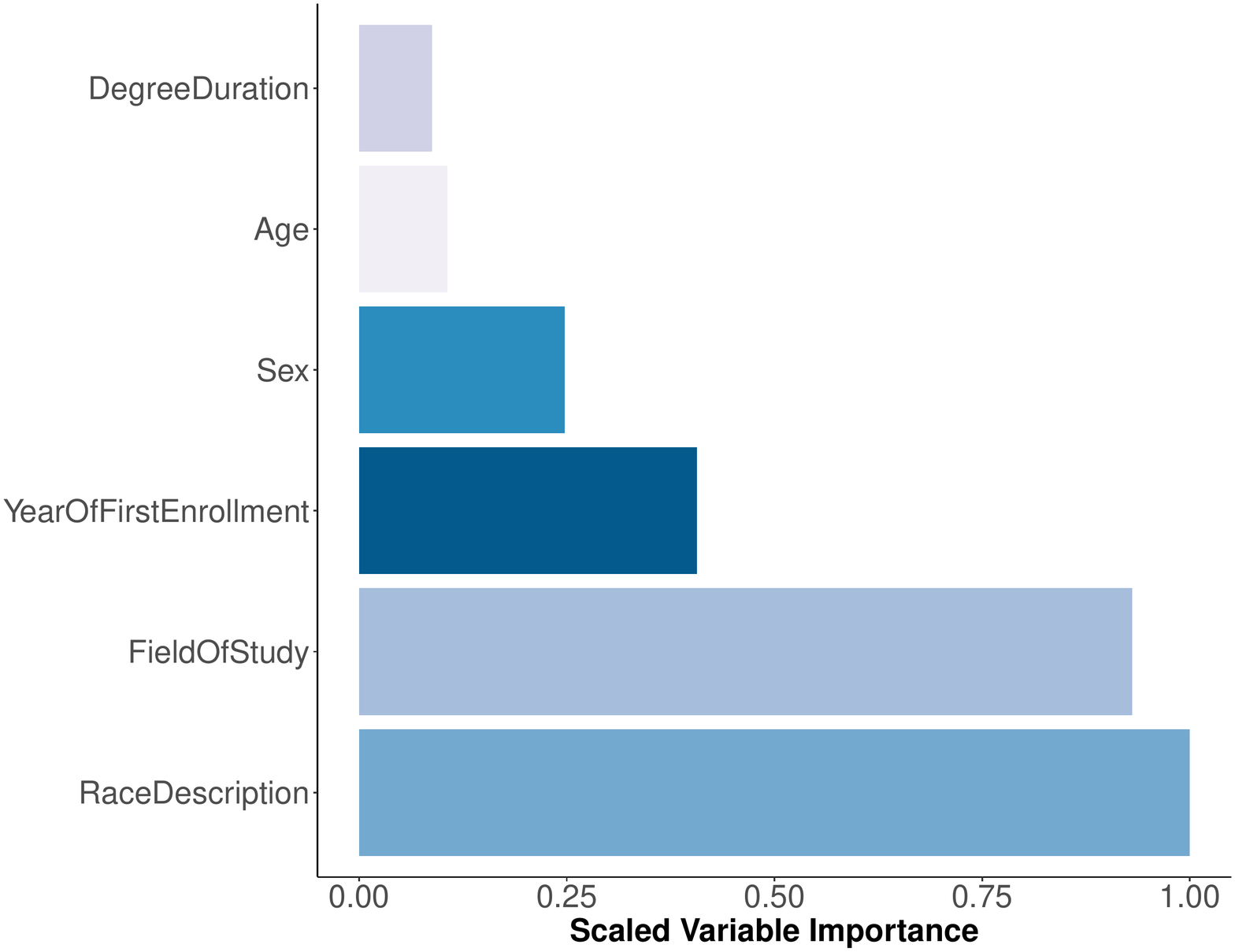}
  \label{fig:tp_varimp_n}
  }
  \subfigure[Variable importance for $N+1$ years]{
  \includegraphics[width=0.47\textwidth]{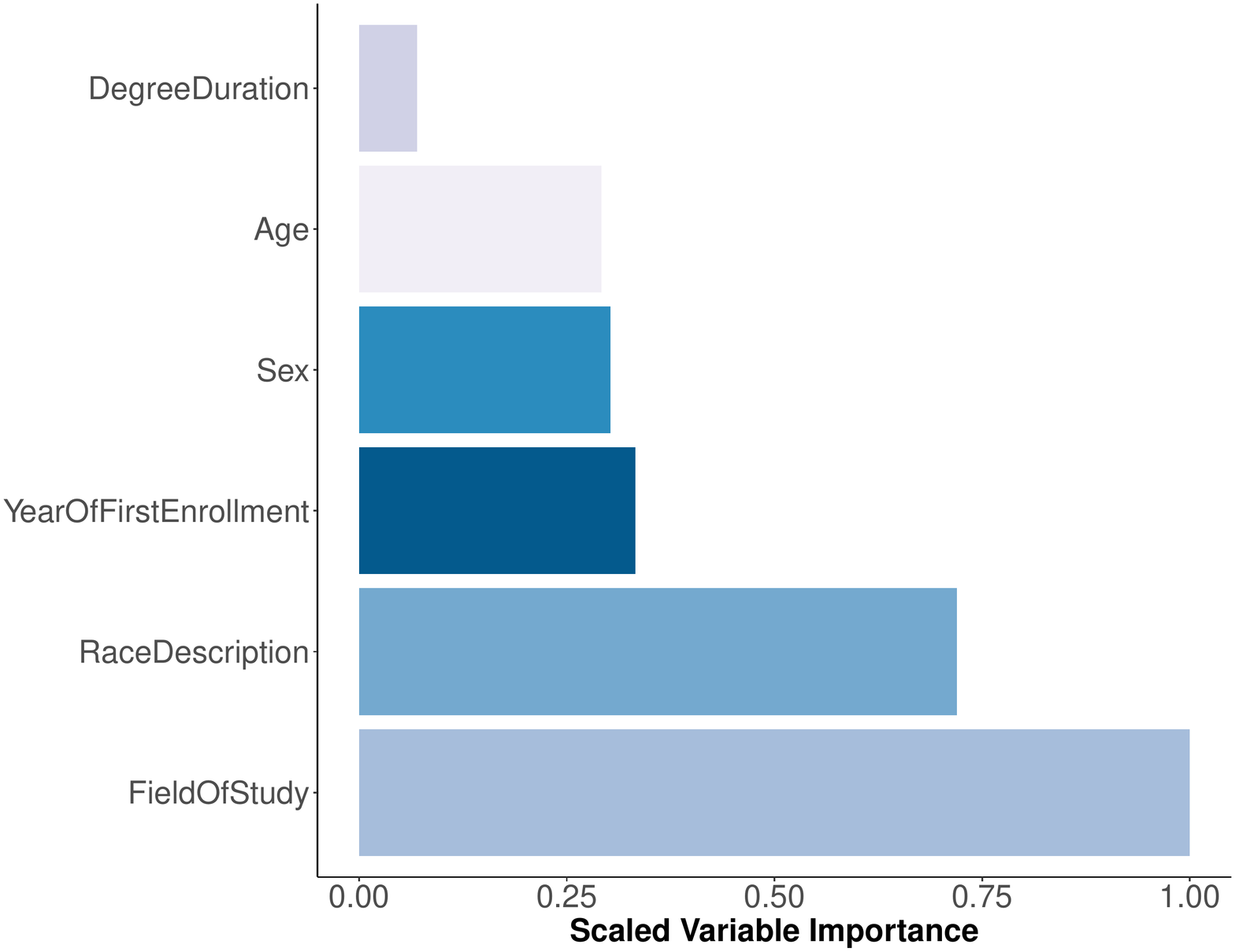}
  \label{fig:tp_varimp_n1}}
  \subfigure[Variable importance for $N+2$ years]{
  \includegraphics[width=0.47\textwidth]{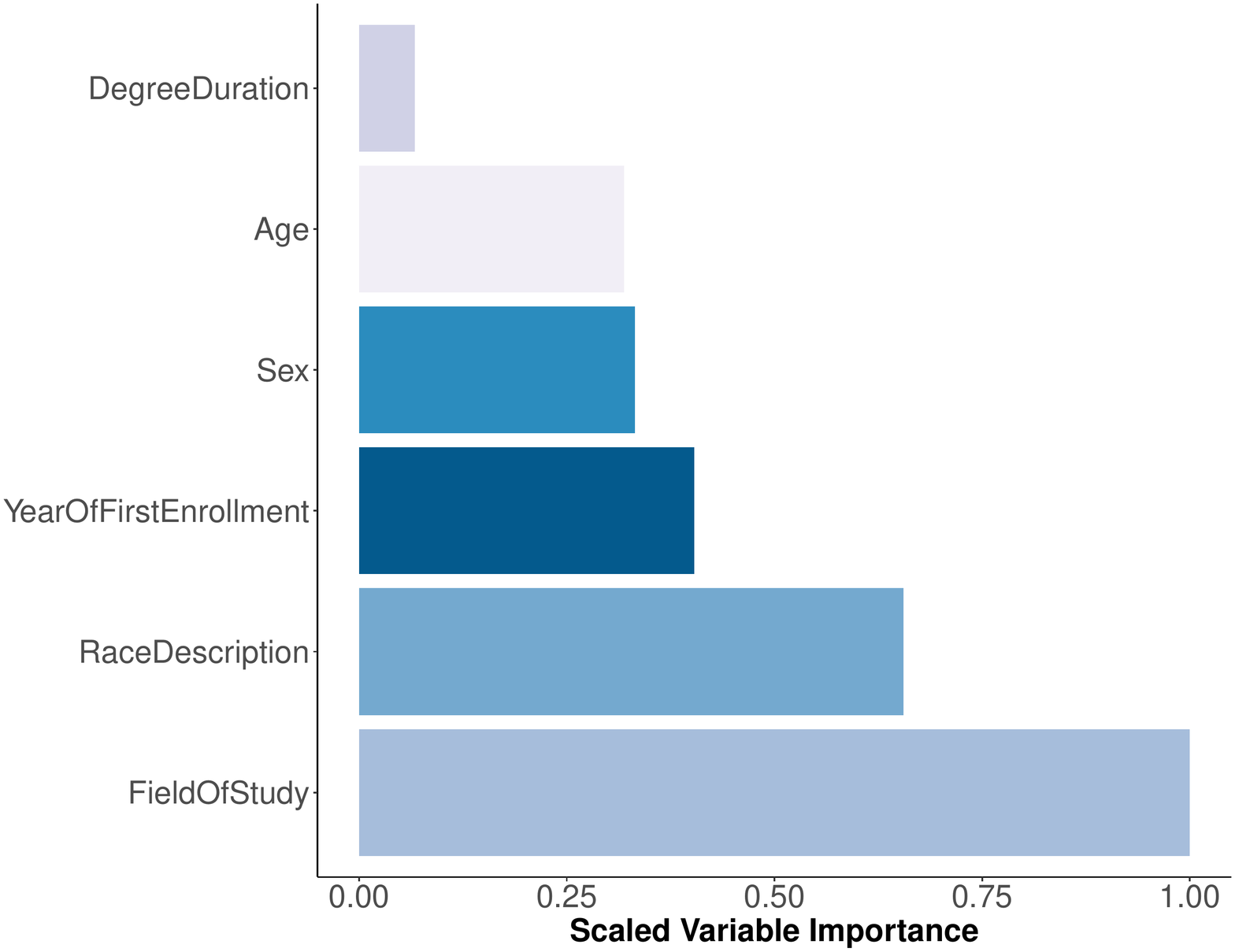}
  \label{fig:tp_varimp_n2}
}
 
\caption{Scaled variable importance results from the different  gradient boosting models. Figure \ref{fig:vip}\subref{fig:tp_varimp_n} shows variable importance for degree completion in the minim required $N$ years. Figures \ref{fig:vip}\subref{fig:tp_varimp_n1} and \ref{fig:vip}\subref{fig:tp_varimp_n2} depict scaled variable importance measure for $N+1$ and $N+2$ completion respectively.      }
\label{fig:vip}
\end{figure*}

\begin{figure*}[ht]
\centering
  \subfigure[Throughput by year of first enrollment]{
  \includegraphics[width=0.45\textwidth]{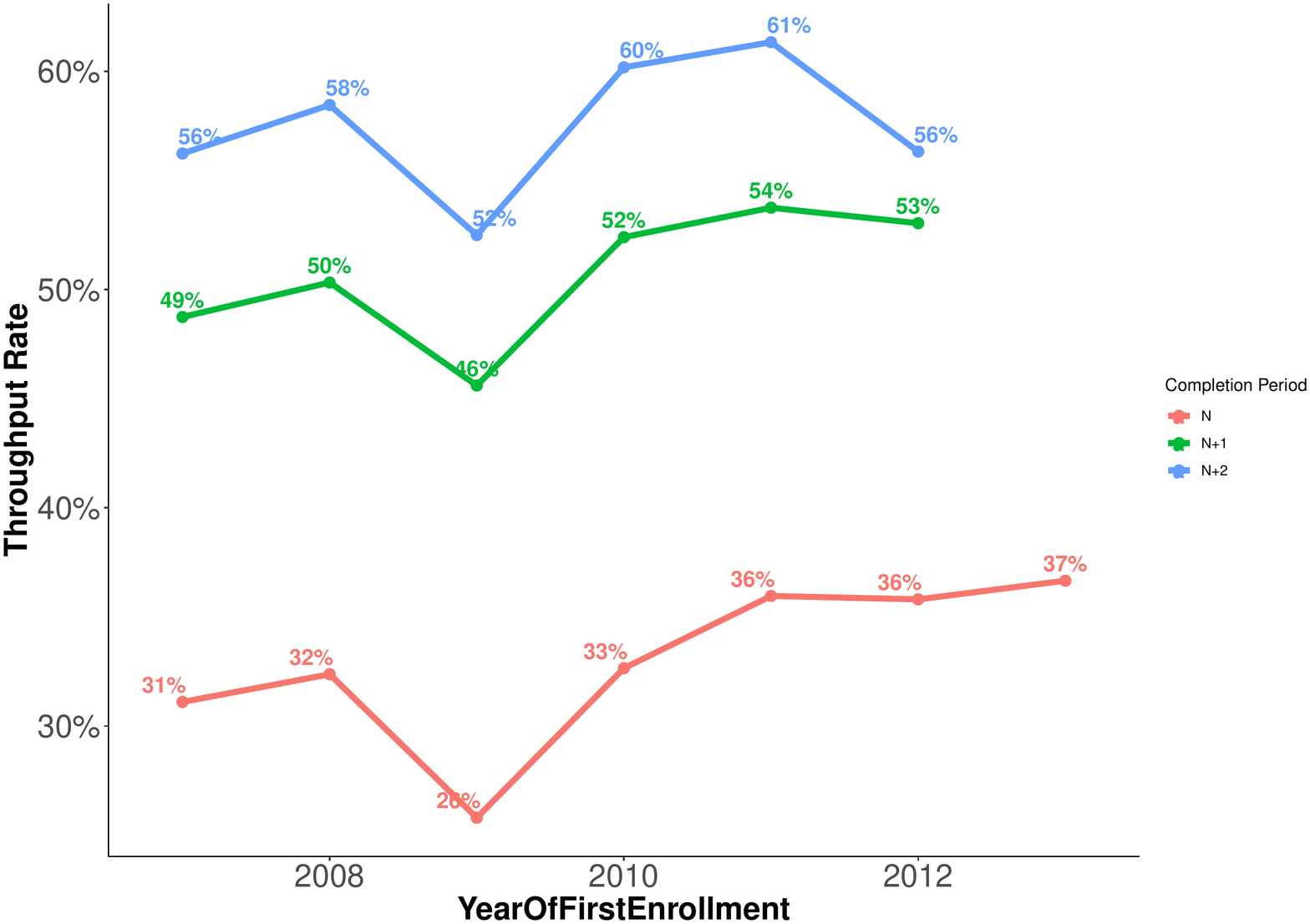}
  \label{fig:tp_year}
  }
   \subfigure[Throughput by race description]{
  \includegraphics[width=0.45\textwidth]{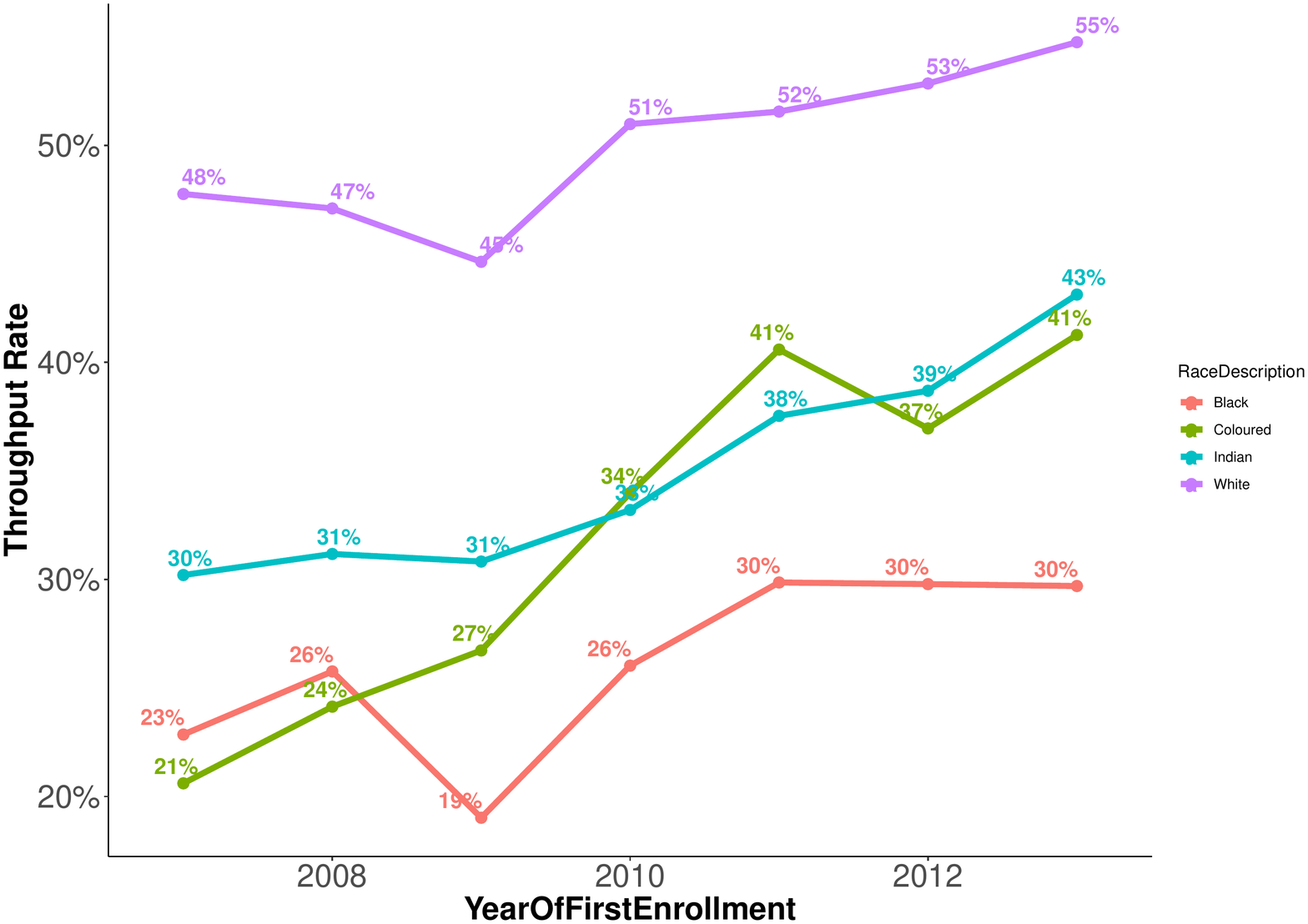}
  \label{fig:tp_race}
  }
  \subfigure[Throughput by sex]{
  \includegraphics[width=0.45\textwidth]{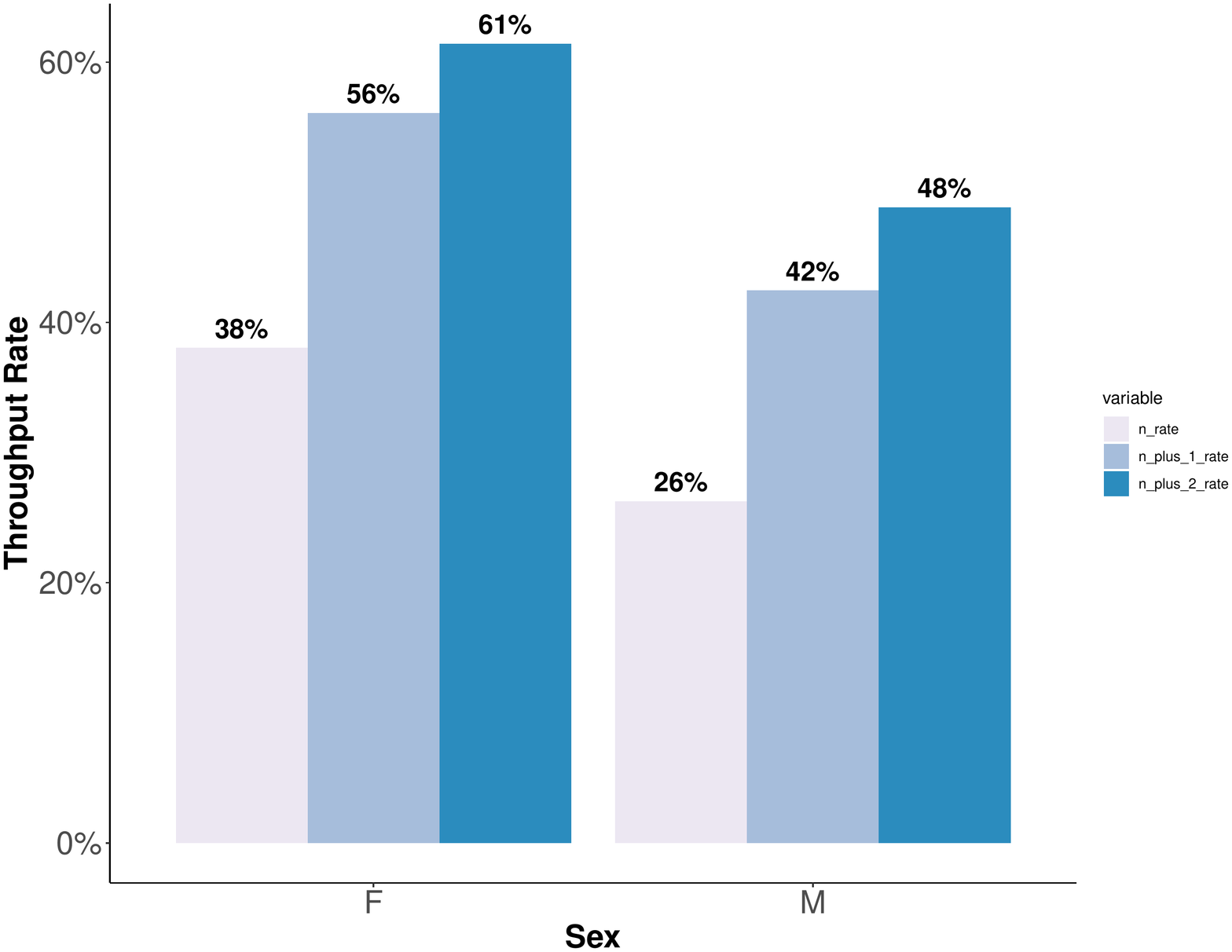}
  \label{fig:tp_sex}}
  \subfigure[Throughput by Field of Study]{
  \includegraphics[width=0.45\textwidth]{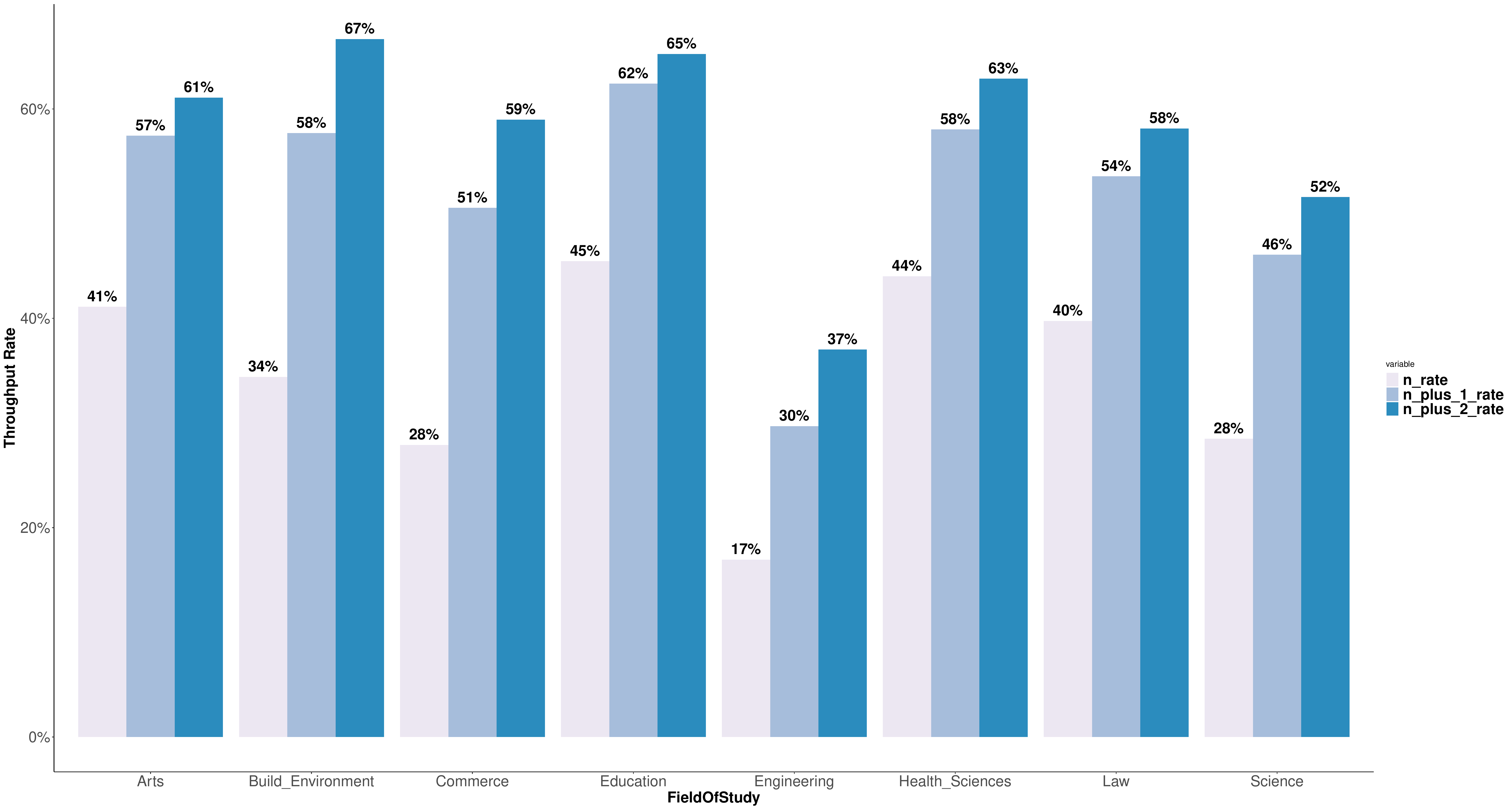}
  \label{fig:tp_fac}
}
 
\caption{Plots illustrating empirical throughput rates by year of first enrollment \ref{fig:emp}\subref{fig:tp_year}, race description \ref{fig:emp}\subref{fig:tp_race}, sex \ref{fig:emp}\subref{fig:tp_sex} and field of study \ref{fig:emp}\subref{fig:tp_fac}.   }
\label{fig:emp}
\end{figure*}

Figure \ref{fig:vip} shows the gradient  boosting scaled variable importance statistics  for each of the three models predicting degree completion time. For completion in the required minimum time $N$ years; race description, field of study, sex and year of first enrollment are identified as the most important predictors. The high importance of the field of study variable suggest differential success rates between faculties. This effect is also observed in figure \ref{fig:emp}\subref{fig:tp_fac} in which the field of study of engineering displays significantly lower throughput rates compared to other faculties. Success differentials by sex are reflected in figure \ref{fig:emp}\subref{fig:tp_sex} where female students display higher throughput rates. The year of first enrollment is further identified as an important predictor, this can interpreted as a cohort effect which can be driven by changes in  both the pre-university and university environments e.g syllabus and admission criteria changes. Figure \ref{fig:emp} shows detailed visual empirical analysis of each the factors.

Figures \ref{fig:vip}\subref{fig:tp_varimp_n1} and  \ref{fig:vip}\subref{fig:tp_varimp_n2} show the scaled variable importance statistics for predicting completion in $N+1$ and $N+2$ years. The order of predictors remains fairly similar to that of predicting completion in $N$. However an important shift to note is the increase in importance of field of study relative to the race description variable. 

\section*{Discussion}\label{discussion}

\subsection*{Model performance}
We have found that both gradient boosting and logistic regression display adequate predictive performance when predicting throughput in N, N+1 and N+2 respectively. The gradient boosting machine displays marginal outperformance in terms of the AUC and AUPR, suggesting that the relationship between throughput and socio-demographic factors is non-linear. This is consistent with the findings of \cite{aulck2017stem}, \cite{huang2011predictive}, and \cite{musso2013predicting}. It is important to note that other explanatory variables more indicative of secondary level attainment could significantly increase predictive power \citep{case2014assessing}. 

\subsection*{Student success and demographics} 
We find that demographic variables, in particular, race, is an important factor in predicting student success, especially when looking at completion on time (N). The race description variable often includes socio-economic factors such as income, quality of secondary schooling, and first generation university entrance, which are reflective of the inequalities within the South African society \citep{parker2006effect}. Similar studies, such as \cite{van2017higher}, \cite{perry2015analysis}, and \cite{strugnell2017throughput} have found race to be a strong determinant of both opportunity and success at university in South Africa.

A key policy signal to counteract this effect is demonstrated by the reduction in importance of race when looking at completion in N+1 and N+2. This illustrates that racial disparities can be addressed through interventions that aim to bridge the articulation gap for students from educationally disadvantaged backgrounds. Parameters that define the articulation gap include university preparedness in both the academic and social aspects, as well as financial pressures that Black African students face \citep{Lewin2014student}. These students, who are usually first generation, arrive at university from poorly resourced schools, without a social network of support, and little financial security. \cite{strugnell2017throughput} have found interventions such as financial support in the form of student loans, bursaries and scholarships; academic support, psycho-social support, and extended degree programmes as effective for reducing this gap. Further evidence suggests that academic literacy modules, which seek to enhance reading and writing skills in different disciplines are effective in this regard \citep{van2008acquiring}. These interventions do not only serve to improve the efficiency of the higher education system, but are imperative for the reduction of poverty and inequality in South Africa.  

Additionally, there has not been sufficient change in teaching practices from those that were mainly targeted at students from predominantly homogeneous backgrounds to accommodate diversity in culture and language that is reflective of a post democratic South Africa \citep{van2008acquiring}. \cite{strydom2014student} suggest using student engagement data from the national South African Survey of Student Engagement to inform teaching practices that will allow for inclusion and consequently better student outcomes across socio-demographic lines.  

\subsection*{The STEM challenge}
The results suggest that field of study is of high importance when predicting throughput and student success. The empirical data analysis in figure \ref{fig:emp} further shows that within faculties, Engineering and Science yield low throughput rates when compared to other disciplines. Low throughput and completion rates in these areas exacerbate the  national and global skills deficit in these critical disciplines which necessitates urgent and high impact interventions. \cite{aulck2017stem} found interventions such as additional mathematics education and entry level introductory courses to be positively correlated with performance in STEM.

\section*{Conclusions}
We have developed an ensemble tree based model for predicting degree completion at a South African University. We used this model to identify significant socio-demographic factors for degree attainment in specified periods. The results show that the models display adequate predictive performance in terms of both precision and sensitivity. The results further indicate that socioeconomic proxies such as race, are highly correlated to student success, with this effect diminishing as the time to completion increases. The results also highlight challenges with throughput in STEM disciplines. We make recommendations on interventions which include extended degree programmes, academic, psychosocial and financial support. 

\FloatBarrier
\bibliography{bib}

\end{document}